# Common neighbours and the local-community-paradigm for topological link prediction in bipartite networks

**Authors:** Simone Daminelli[2,†], Josephine Maria Thomas[1,†], Claudio Durán[1,3] & Carlo Vittorio Cannistraci[1,*]

**Affiliations:**

[1]Biomedical Cybernetics Group, Biotechnology Center (BIOTEC), Technische Universität Dresden, Dresden, Germany

[2]Bioinformatics Group, Biotechnology Center (BIOTEC), Technische Universität Dresden, Dresden, Germany

[3]Escuela de Ingeniería en Bioinformática, Universidad de Talca, 2 Norte #685, 3465548, Talca, Chile

*To whom correspondence should be addressed: kalokagathos.agon@gmail.com

†These authors contributed equally to this work

**Subject Areas:**

Complex systems and networks

Interdisciplinary physics

Biological physics

Link prediction

Bipartite networks

Network models

**ABSTRACT**
Bipartite networks are powerful descriptions of complex systems characterized by two different classes of nodes and connections allowed only across but not within the two classes. Unveiling physical principles, building theories and suggesting physical models to predict bipartite links such as product-consumer connections in recommendation systems or drug-target interactions in molecular networks can provide priceless information to improve e-commerce or to accelerate pharmaceutical research. The prediction of nonobserved connections starting from those already present in the topology of a network is known as the link-prediction problem. It represents an important subject both in many-body interaction theory in physics and in new algorithms for applied tools in computer science. The rationale is that the existing connectivity structure of a network can suggest where new connections can appear with higher likelihood in an evolving network, or where nonobserved connections are missing in a partially known network.

Surprisingly, current complex network theory presents a theoretical bottle-neck: a general framework for local-based link prediction directly in the bipartite domain is missing. Here, we overcome this theoretical obstacle and present a formal definition of common neighbour index (CN) and local-community-paradigm (LCP) for bipartite networks. As a consequence, we are able to introduce the first node-neighbourhood-based and LCP-based models for topological link prediction that utilize the bipartite domain. We performed link prediction evaluations in several networks of different size and of disparate origin, including technological, social and biological systems. Our models significantly improve topological prediction in many bipartite networks because they exploit local physical driving-forces that participate in the formation and organization of many real-world bipartite networks. Furthermore, we present a local-based formalism that allows to intuitively implement neighbourhood-based link prediction entirely in the bipartite domain.

**INTRODUCTION**
Network science is a field that is gaining increasing interest in many scientific communities and in particular among physicists that model the organization of complex systems. The organization of real networks is usually driven by both mechanistic (or pattern-based) factors that are predictable and random factors that are unpredictable; hence, in principle, only the former can be explained using physical modelling (Lü et al. 2015). In the challenging quest for the principles, the mechanisms and the hidden driving-forces that shape real-world networks, a milestone was set by a recent study where physicists suggested that the extent to which the formation of a network can be explained coincides with our ability to predict missing links (Lü et al. 2015). This study gave a significant momentum to the theory of link prediction based merely on the connectivity structure of a network – a.k.a. topological link prediction – which till now was mostly perceived as an applied problem in monopartite networks (Liben-Nowell & Kleinberg 2007). In fact, given a monopartite undirected and unweighted network $G(V,E)$ where $V$ is the set of observed nodes and $E$ is the set of observed links, the topological link prediction problem consists in estimating the likelihood that each nonobserved link (missing link) between the observed nodes exists. The ranking of the nonobserved links in order of decreasing likelihood represents a list of candidate links to be employed for modelling the network connectivity growth. Indeed, the extent to which the process of network formation is explicable by a method coincides with the method capability to predict missing links (Lü et al. 2015). There are two main families of methods: model-based and model-learning. The first approach is assuming that a well-characterized physical mechanism is the primary driving force behind the network organization (Lü et al. 2015), and is based on an explicit deterministic model that simulates such a mechanism. The second type does not make a priori assumptions about any specific organization principle of the network topology, and instead relies on an implicit stochastic model-learning: providing at each step a different solution that should converge to the hidden network behaviour in a sufficient number of iterations.

Elaborate methods, often with several parameters to tune, have been introduced in the class of stochastic models (Cannistraci et al. 2013). Good performance has been shown by stochastic block models also considering weighted bipartite networks (Guimerà et al. 2012), and by specific methods exploiting even the topological information related with temporal dimension of evolving networks (Dunlavy et al. 2011). However, link prediction based on network edge weights or on multiple time-based configurations of the same network is not in the scope of this article, because the majority of these gracefully designed techniques rely on more specific features (not always available for all the types of networks) or present some major drawbacks (Cannistraci et al. 2013). For instance, apart from the problem of tuning the model parameters in an unsupervised framework, a severe limitation is their high computational time, which in practice reduces their immediate application to networks of small dimensions (no more than few hundreds of nodes), in comparison to the large networks used in real problems. For these reasons, methods based on explicit models represent an efficient and parameter-free alternative widely employed in many applicative domains, and the node-neighbourhood-based (local-based) indices are the most relevant among them. The rationale behind these methods is that the likelihood of an interaction between two non-adjacent nodes is strongly related with mechanisms of organization involving their first and/or second neighbour nodes. The common neighbours (CN) index, which is the precursor of these methods, follows the intuition that the likelihood that two seed nodes x and y interact increases if their sets of first-node-neighbours overlap substantially: this implies that the larger the number of common neighbours, the higher the likelihood that the two seed nodes interact. The network organization model behind the CN index is named *triadic closure* (Opsahl 2013) (Fig.1A), which is a basic generating mechanism of communities in complex networks (Bianconi et al. 2014). Many other node-neighbourhood-based indices can be interpreted as a variation or generalisation of CN: Jaccard's index (JC) is a normalisation of CN; Adamic & Adar (AA) and Resource Allocation (RA) indices give more importance to CNs with low degree (Cannistraci et al. 2013). CN and all derived measures following the triadic closure principle are very powerful topology-based methods for link prediction in many monopartite undirected and unweighted real networks from various fields (Liben-Nowell & Kleinberg 2007). However, it has been shown that due to bipartite networks' specific properties (Opsahl 2013) all topological methods based on the triangle closing model cannot work in the bipartite case (Kunegis et al. 2010; Kunegis 2014). Thus, at the moment, both CN index and all the related variations (JC, AA and RA) are not defined for bipartite networks. On the other hand, out of the node-neighbourhood-based link prediction methods, the preferential attachment (Barabási & Albert 1999; Newman 2001) (PA) model can be used in bipartite networks, and has been shown to perform better than various algebraic (e.g. matrix factorization) methods in many bipartite real networks (Kunegis et al. 2010).

In contrast to the existing node-neighbourhood-based approaches, a new strategic shift has been introduced recently in which the focus is no longer only on groups of common nodes and their node neighbours, but also on the organization of the links between them (Cannistraci et al. 2013). This theory, defined and tested only in monopartite undirected and unweighted networks, is known as the local community paradigm (LCP-theory) (Cannistraci et al. 2013). In the past, approaches inspired by a similar rationale have been introduced by Clauset in the context of community detection (Clauset 2005). The LCP-theory (Cannistraci et al. 2013) holds that for modelling link prediction in complex networks, the information content related with the common neighbour nodes should be complemented with the topological information emerging from the interactions between them. The cohort of common neighbours and their cross-interactions form what is called a local community; the cross-interactions between CNs are called local community links (Fig. 1A). In order to demonstrate the validity of the theory on several classes of networks, different classical node-based link prediction techniques like CN, JC, AA, RA and PA were reinterpreted according to the LCP-theory (Cannistraci et al. 2013), by introducing terms related with the local community links (LCLs) in their formulations. This mathematical reformulation represents the Cannistraci variations of CN, JC, AA, RA and PA respectively renamed CAR, CJC, CAA, CRA and CPA (Cannistraci et al. 2013). For simplicity, from here on we will refer to the former as *classical models* and the latter as *LCP-based models*. All the details on these models are provided in the Methods.

A theoretical innovation of this article is the definition of the LCP-theory and the relative local-based models for link prediction in undirected and unweighted *bipartite* networks. However, the first and nontrivial step required for this extension is the definition of the concept of CN index in bipartite topologies: surprisingly, as mentioned above, a concept not yet formally defined in network theory.

## METHODS
### Definition of CNs and LCP-theory in bipartite networks

The definition of CNs in *monopartite* networks is related to the concept of triadic closure. In an undirected and unweighted monopartite graph, the triadic closure produces a triangle when an edge connects two nonadjacent nodes which already have a CN. Therefore, any node connected to two nonadjacent nodes that might be involved in the triangular closure between them is a CN. Moreover, the triadic closure has the main property of enclosing the shortest path between the two nonadjacent nodes. Thus, intuitively the definition of CNs in bipartite networks turns out to be related to quadrangular closure (Fig. 1B) that is the equivalent of triangular closure operation in bipartite topologies (Latapy et al. 2008). In fact, in bipartite structures, the quadrangular closure encloses the shortest path between two non-adjacent nodes that belong to two distinct classes (Fig. 1B). This entails that in bipartite networks we define the CNs of two given nonadjacent seed nodes (for which we want to estimate the likelihood of their missing interaction) as the nodes that are involved in all possible quadrangular closures between these seed nodes, and the LCLs as all the links that occur between these CNs (Fig. 1B). For the sake of clarity, we notice that the Kunegis definition of the P3 link prediction index in bipartite networks (Kunegis 2014) counts the number of paths of length three between two seed nodes. Thus, it is mathematically (not conceptually) equivalent to our LCLs definition, but not to our CNs definition. For instance, in Fig. 1B the first neighbourhood of the two seed nodes $x$ and $y$ counts 6 CNs and 7 LCLs. Analogously, for two given adjacent seed nodes for which we want to estimate the reliability of their existing interaction, we define the CNs as the nodes that are involved in all already existing quadrangles passing through the seed nodes, and the LCLs as all the links that occur between these CNs. A practical consequence of these definitions is that mathematically, both the CN, JC, AA and RA indices and their Cannistraci variations for prediction in the bipartite networks, can be expressed using the same formula provided for monopartite networks.

***Classical node-neighbourhood-based* and *LCP-based* models for link prediction.***
Given two non-adjacent nodes x and y, the classical formulations (Cannistraci et al. 2013) are:

$$CN(x,y) = |N(x) \cap N(y)|$$

$$JC(x,y) = \frac{CN(x,y)}{|N(x) \cup N(y)|}$$

$$AA(x,y) = \sum_{s \in N(x) \cap N(y)} \frac{1}{log_2|N(s)|}$$

$$RA(x,y) = \sum_{s \in N(x) \cap N(y)} \frac{1}{|N(s)|}$$

$$PA(x,y) = |N(x)| \cdot |N(y)|$$

where $N(x)$ is the set of neighbours of node $x$ and $|N(x)|$ its cardinality or degree.
Whereas the Cannistraci variations (Cannistraci et al. 2013) of the same indices are:

$$CAR(x,y) = CN(x,y) \cdot LCL(x,y);$$

$$CJC(x,y) = \frac{CAR(x,y)}{|N(x) \cup N(y)|}$$

$$CAA(x,y) = \sum_{s \in N(x) \cap N(y)} \frac{|\gamma(s)|}{log_2|N(s)|}$$

$$CRA(x,y) = \sum_{s \in N(x) \cap N(y)} \frac{|\gamma(s)|}{|N(s)|}$$

where: *LCL* is the number of local community links (see Fig. 1B); the node *s* refers to one of the common neighbours; $\gamma(s)$ refers to the sub-set of neighbours of *s* that are also common neighbours of x and y, thus $|\gamma(s)|$ is the local community degree of *s* and corresponds to the number of local community links that originate from *s*.

For PA, a different mathematical reformulation of this index in function of CN was proposed (Cannistraci et al. 2013):

$$PA(x,y) = |N(x)| \cdot |N(y)| = [e(x) + CN(x,y)] \cdot [e(y) + CN(x,y)]$$
$$= e(x) \cdot e(y) + e(x) \cdot CN(x,y) + e(y) \cdot CN(x,y) + CN(x,y)^2$$

$e(x)$ is the external degree of *x* (Fig.1) computed considering the neighbours of *x* that are not common neighbours. Thus, the Cannistraci variation is:

$$CPA(x,y) = e(x) \cdot e(y) + e(x) \cdot CAR(x,y) + e(y) \cdot CAR(x,y) + CAR(x,y)^2$$

As a further remark, PA (Kunegis et al. 2010) and P3 (Kunegis 2014), the second of which was discussed in the first section of the Methods, are the only node-neighbourhood-based local models that have already been applied for link prediction in bipartite networks (Kunegis 2014). All the other node-neighbourhood-based and LCP-based indices for link prediction in bipartite networks are introduced here for the first time. As a further remark, all these algorithms are parameter-free, which is an important advantage for real applications.

**State-of-the-art link prediction methods based on the one-mode-network projections.**
It has been shown that any bipartite network can be projected into its two monopartite network representations (called one-mode-projection) by means of a procedure called bipartite network projection (BNP) (Zhou et al. 2007). Given a bipartite network characterized by two classes of nodes A and B, prediction of links between the two classes is possible through the vector representation of the two one-mode projections (Yildirim & Coscia 2014). For instance, if A={a1, a2, a3} and B={b1, b2} are the nodes from the two classes A and B present in the bipartite network, a vector b1 = (0, 1, 0) indicates that node b1 interacts with node a2 only. Then, the missing links b1-a1 and b1-a3 can take likelihood scores based on how similar (according to a certain metric) the vector representations of a1 and a3 are to the a2 vector, which is the node interacting with b1 (Yildirim & Coscia 2014). Unfortunately, one-mode projections can cause loss of the original topological information present in the bipartite network structure (Larremore et al. 2014).

Here, we selected the following advanced metrics, which represent the state of the art for link prediction in bipartite networks based on the two one-mode projections (the technical details of the respective procedures are provided in the cited references): Network Based Inference (NBI, also known as ProbS) (Zhou et al. 2007), Bipartite Projection via Random-walk (BPR) (Yildirim & Coscia 2014); and baseline spatial distance measures: Jaccard (Jac), Euclidean (Euc), Cosine (Cos) and Pearson (Pea) (Yildirim & Coscia 2014).

All these measures have been computed as described by Coscia *et al.* (Yildirim & Coscia 2014), using a python implementation provided by the authors: (http://www.michelecoscia.com/?page_id=734).

Subsequently to the NBI (or ProbS), the HeatS and a method based on the hybrid of the two (named HeatS+ProbS) were also proposed by the same authors of NBI (Zhou et al. 2010). In particular, the hybrid method showed some improvements, but this gain of performance was paid by including a free parameter $\lambda$ into the hybrid model, whereas NBI is a tune-free algorithm (i.e., does not depend on any control parameters), which is - in the words of the authors of these two methods - "a big advantage for potential users" (Zhou et al. 2007). Furthermore, a study by Liu and Zhou (Liu & Zhou 2012) investigated the effect of heterogeneity in initial resource configurations on the HeatS+ProbS hybrid algorithm and discovered that both recommendation accuracy and diversity could be improved. Unfortunately, this enhancement (that was proven on only two benchmark datasets) bore the cost of including also a second parameter $\eta$ in the HeatS+ProbS algorithm, which is - according to the authors of the article - a limitation for the application of the proposed method (Liu & Zhou 2012). To conclude, in our comparisons we decided to adopt NBI because it represents an important tune-free algorithm, which was a reference for all the aforementioned methods. On the other hand, we decided to also adopt BPR as an additional parameter-free algorithm, which was shown by Yildirim and Coscia to have a comparable performance for link prediction in various bipartite networks against both NBI and the HeatS+ProbS algorithm (Yildirim & Coscia 2014).

**Bipartite Local Model (BLM) for drug-target prediction**
The Bipartite Local Model (BLM) is a supervised inference method to predict novel drug-target interactions (Yamanishi et al. 2008; Bleakley & Yamanishi 2009). Local models transform the problem of interactions prediction into a binary classification problem, exploiting support vector machine (SVM) and a kernel representation of the interactions in the bipartite domain. Moreover, biological background knowledge (chemical similarity and protein sequence similarity) is given to the classifier to learn a local model and calculate the likelihood of a certain link, given the similarity to the local neighbours of the specific drug (or target) under consideration. The combination of both perspectives gives the overall drug-target interaction ranking and classification.
We used the MATLAB implementation of BLM as provided by the authors (Yamanishi et al. 2008; Bleakley & Yamanishi 2009) at this link: http://cbio.ensmp.fr/~yyamanishi/bipartitelocal/

**Data description**
We performed link prediction evaluations in six bipartite networks of different size and origin, including technological, social and biological systems. Directed links, if present, were treated as undirected. Weighted links, if present, were treated as unweighted.
1. Aid (Coscia et al. 2013): composed of 151×34 nodes and 1889 interactions. International aid organizations connected to development issues.
2. Ipums (Ruggles et al. 2010): composed of 267×513 nodes and 18088 interactions. Industries connected to the fields of education of people they employ.
3. Movielens100k (http://grouplens.org/datasets/movielens/): composed of 1682×943 nodes and 100.000 interactions. It contains movie ratings from http://movielens.umn.edu/. Left nodes are users and right nodes are movies. An edge between a user and a movie indicates that the user has rated the movie.

4. G-Protein Coupled Receptors (GPC Receptors) (Yamanishi et al. 2008): composed of 95×223 nodes and 635 interactions. Drugs binding G-protein coupled receptors.

5. Ion Channels (Yamanishi et al. 2008): composed of 204×210 nodes and 1476 interactions. Drugs binding ion channel proteins.

6. Enzymes (Yamanishi et al. 2008): composed of 664×445 nodes and 2926 interactions. Drugs binding enzyme proteins.

Basic topological statistical features are reported in Suppl. Table S1.

**RESULTS AND DISCUSSION**

Comparing the proposed LCP-based and classical models (which work in the bipartite domain) to the state-of-the-art methods (which are based on the one-mode-network projections) we found that in general the LCP-based link predictors offer a significant improvement in many technological, social and biological (molecular drug-target) bipartite networks of assorted size. Suppl. Fig. 1 shows the precision of each tested link predictor in recovering L edges, where L is the number of edges equivalent to 10% of the original edges present in the network. For each network L edges were randomly removed and the remaining observed network topology was used to apply the link predictor, which offered as output the list of candidate links (composed of the nonobserved and the removed edges) ranked in decreasing likelihood of existence. The precision represents the ratio of correct edges recovered out of the top L edges in the candidate list generated by each link predictor. This operation was repeated 100 times for each network and we report the mean and standard error for each method in Suppl. Fig.1. The described procedure is the gold standard for quantification of link prediction performance (Lü et al. 2015). In Fig.2, where the LCP-based models demonstrate again clear superiority, we provide the values obtained using the same procedure for estimation of another measure called area under precision recall curve (AUPR). While precision assesses the ability of each method to rank the L removed edges among the top L in the candidate list, AUPR assesses the ability to rank the L removed edges the highest possible in the entire candidate list. For this reason AUPR is considered a more complete and robust estimation of performance (Yang et al. 2014). In contrast, the area under the ROC curve has recently been proved to be very deceptive in comparison to AUPR (Yang et al. 2014), and it should be avoided in evaluation of link prediction. Furthermore, Fig. 3A-B, shows the mean precision and mean AUPR across all networks. Taken together, these results emphasize that LCP-based models can significantly (p-value <0.001, Fig.4) overcome - at least in bipartite networks topologically similar to the ones considered here - state of the art methods for link prediction, offering 123% improvement in precision compared to classical models, and 186% improvement in precision compared to one-mode-projection methods (Fig.4A,C). The performance improvements evaluated according to AUPR are even larger (Fig.4B,D).

Traditionally, prediction of drug-target interactions using three-dimensional molecular structure information is a hot topic in biophysics. Molecular docking is one of the preferred modelling approaches, however, it is only applicable when the ligand is a relatively small molecule in comparison to the receptor and/or experimental data on the binding site are accessible (Valencia & Pazos 2002). In addition, molecular dynamics simulations to further analyse obtained docking solutions are often needed in order to calculate the affinity between two molecules, which can be computationally expensive if a rigorous exploration of a large conformational space is required (Valencia & Pazos 2002). Recently, it has been proven that building machine learning classifiers that integrate prior topological systemic knowledge (from the drug-target network connectivity) with chemical and sequence similarity knowledge (from the single molecules that constitute the nodes of such molecular networks) is effective in predicting multiple new drug-target interactions (Yamanishi et al. 2008; Bleakley & Yamanishi 2009). Thus, exploiting exclusively the network topology to predict the likelihood of molecular interactions is an interesting interdisciplinary topic in physics: a proof of concept that can pave the way to build in the future hybrid methods where network modelling (at systemic level) is used to infer prior information useful to support predictions at molecular structural level.

In the three drug-target networks here analysed (Fig.2D-F), we can ascertain that LCP-based models provide noteworthy link prediction performance. This is demonstrated by comparing the performance of LCP-based models versus a *supervised* model named Bipartite Local Model (BLM), which is considered a baseline algorithm specifically developed for prediction of drug-target interactions in bipartite molecular networks (Yamanishi et al. 2008; Bleakley & Yamanishi 2009). BLM builds a classifier using not only specific local network topology information of each drug or target directly in the bipartite domain, but also integrating prior knowledge (chemical and sequence similarities) between the molecules that constitute the nodes of such biological networks (Yamanishi et al. 2008; Bleakley & Yamanishi 2009). We should expect that the performance of BLM is significantly higher than that of LCP-based models, which are unsupervised and exploit only the topological bipartite information. However, we unexpectedly observed that, in two out of three tested networks, the best of LCP-based models are not significantly different and sometimes are even better than BLM (Fig. 3C,D and Suppl. Fig. 2-3).

Hitherto, we have given a practical demonstration that our proposed LCP-theory works because the methods based on it provide significant improvements in link prediction performance. However, we did not offer any explicit and convincing explanation as to why the LCP-theory can boost link prediction in many networks of such different origin. In order to understand from the physical point of view why LCP-based models are so effective, we should directly examine the structure of such bipartite networks. Fig. 5 displays the local-community-paradigm decomposition plot (LCP-DP) (Cannistraci et al. 2013) for each of the analysed networks. The LCP-DP is a form of network-decomposition because each existing link in the network is plotted in a bi-dimensional space according to both its number of CNs (on the x-axis) and the respective number of LCLs (number of links between the CNs, on the y-axis). The result of this decomposition is a plot that offers a link-based visualisation of the structure of the analysed network and provides information on the presence and size of its local communities.

A main discovery, which emerges from the LCP-DP, is that the structure of all the analysed bipartite networks present a common pattern: a strong correlation between the two variables CN and LCL, which we called the LCP-correlation (LCP-corr) (Cannistraci et al. 2013). This structural organization that complies with the LCP-corr, constitutes the basis of the LCP-theory, and has already been proven for many real-world monopartite networks that are related to dynamic and heterogeneous systems. These systems are characterised by weak interactions that in turn facilitate network adaptability, evolution and remodelling (Cannistraci et al. 2013), which, in fact, are typical features of social and biological systems. Here, we gather that the analysed bipartite networks also have a clear LCP-organization that explains why LCP-based models are so effective. In fact, the likelihood of a non-observed link to exist according to a classical model is high only if (necessary condition) the number of CNs is high. Whereas, according to the LCP-theory a non-observed link has high likelihood only if both CNs and LCLs are simultaneously high. In Fig. 5 the correlation for each network is reported according to Pearson and Spearman formula. The results of Pearson and Spearman correlations are similar for all the networks except for the drug-target networks Ion-Channel and Enzymes. This result denotes structural nonlinearity in the local organization of these networks, which in fact present an irregular LCP-DP that is characterized by important solutions of continuity. The reason for the outlandish structure of these two networks is that they are unsaturated networks with a significant amount of missing information which is 'mirrored' by large holes in the LCP-DP plot. Theoretically, in networks such as these that follow the LCP organization but present significant discontinuity in their topological structure, the LCP-based link predictors should provide an advantage with respect to the classical link predictors based only on the CN information. As a matter of fact, looking at the link prediction results on the Ion Channel (Fig.2E) and Enzymes (Fig.3F) networks, we can notice that the performance of the classical methods is here decreased with respect to the other methods, while the LCP-based link predictors offers always a good performance.

**CONCLUSION**

Here, we introduce a new theory that demonstrates how many bipartite real-world networks have a typical structural organization we called the local-community-paradigm (LCP), which concerns the information derived from the node neighbourhood connectivity (local community) orbiting around each existing link. Current network theory presents a theoretical bottle-neck: a general framework for local-based link-prediction directly in bipartite networks is missing. Hence, we overcome this theoretical obstacle and present a formalism that allows to intuitively realize neighbourhood-based link-prediction entirely in the bipartite domain in networks that are characterized by LCP organization. In fact, the bipartite formulations of the CN index, the classical and LCP-based methods - presented here for the first time - are not only a valuable contribution to improve topological prediction in bipartite complex networks, but also a first effort to intuitively devise local-based link prediction theory completely in the bipartite domain.

The practical implication of our study on link formation prediction in bipartite networks is that the proposed models - which are parameter-free, hence very convenient to apply to concrete problems - can provide, for instance, valuable forecasts to improve e-commerce or to accelerate pharmaceutical research.

**Acknowledgments**

We thank Gregorio Alanis Lobato, Stephan Grill, Sergey Samsonov and Ewa Aurelia Miendlarzewska for the useful suggestions and comments; and for proofreading the article. We kindly acknowledge Michele Coscia for releasing the python code for various bipartite projection methods. We thank Daniel B. Larremore for his recommendations on the bipartite scientific literature.

**Funding**
This work was supported by the independent group leader starting grant of the Technische Universität Dresden (TUD). We acknowledge also support by the Open Access Publication Funds of the TUD, the German Research Foundation (DFG) and the Centre for Information Services and High Performance Computing (ZIH) of the TUD. C.D. was partially supported by the DAAD jung ingenieure programm fellowship +56-2-29462636.

**Author contributions**
C.V.C. and S.D. envisioned the study. C.V.C. and S.D. invented the CN-index and all the related classical node-neighbourhood-based indices for link prediction in bipartite networks. C.V.C invented the LCP theory and the LCP-based models for link prediction in bipartite networks, while S.D. and J.M.T. inspected their correctness. C.V.C. and S.D. designed the experiments and the algorithms. S.D., J.M.T., C.D. implemented and ran the codes and performed the computational analysis with C.V.C. help. All the authors analysed the results. C.V.C. designed the figures and wrote the article with input and corrections from the other authors. J.M.T. and C.D. realized the figures under the C.V.C. guidance. C.V.C. led, directed and supervised the study.

**Competing interests**
The authors declare no competing financial interests.

**Data and materials availability**
The MATLAB code of the link prediction models and the used Networks are available at: https://sites.google.com/site/carlovittoriocannistraci/5-datasets-and-matlab-code

# FIGURES

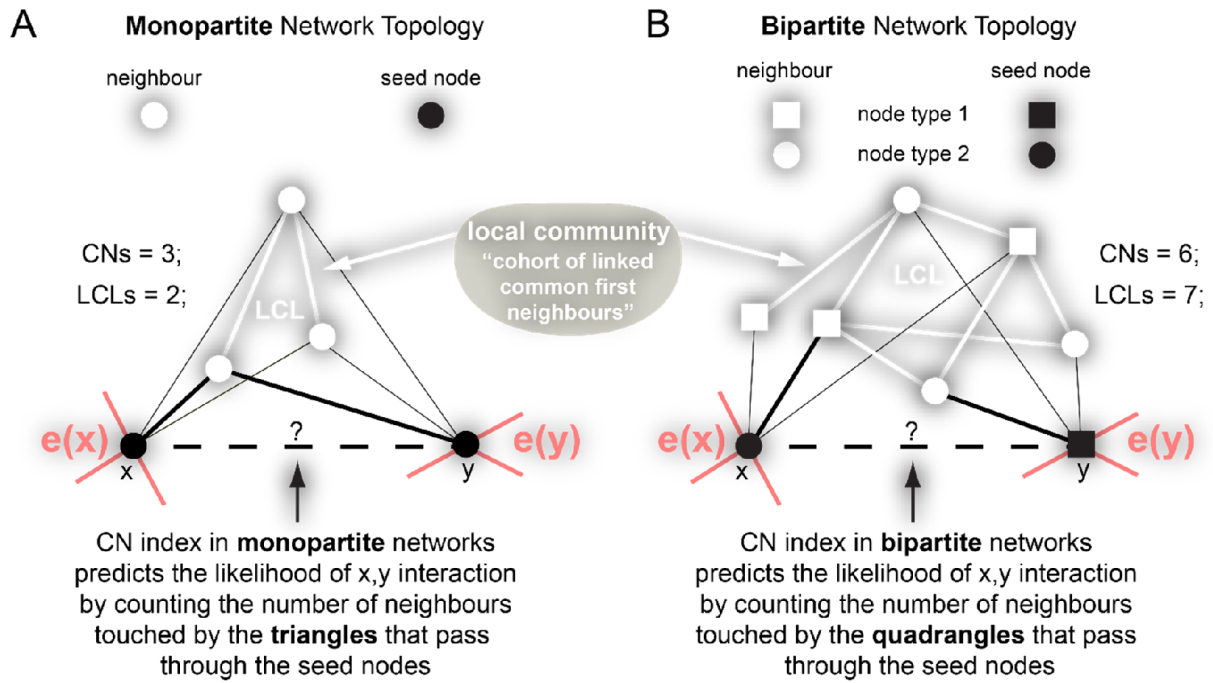

**Figure 1. Definition of common neighbours (CN), local community links (LCL) and local community. (A)** Example of monopartite network topology. CNs are the nodes (white colour) touched by all the possible shortest paths between two nonadjacent nodes, providing **triangular** closure between them. LCLs are the links (white colour) between the CNs. The cohort of CNs and their respective LCLs is named a local community. **(B)** Example of bipartite network topology. CNs are the nodes (white colour) touched by all the possible shortest paths between two nonadjacent nodes, providing **quadrangular** closure between them. LCLs are the links (white colour) between the CNs. The cohort of CNs and their respective LCLs is named a local community.

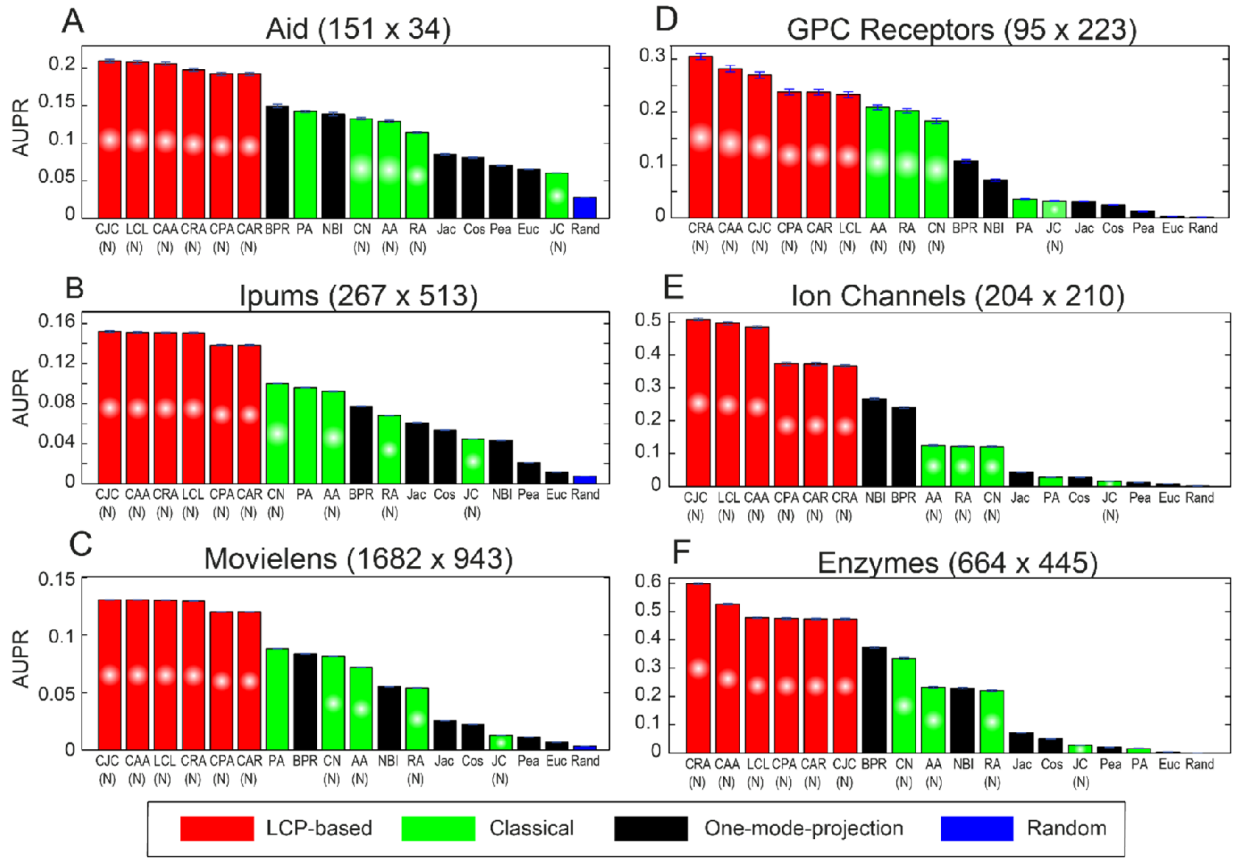

**Figure 2. Link prediction evaluation based on the models' area under the precision recall curve (AUPR).** The AUPR indicates the ability of each model to prioritize the 10% of edges randomly removed from the original network in their candidate list. This operation was repeated 100 times for each network and the mean AUPR and standard error for each method are reported. Colours of the bars define the class of the prediction methods: LCP-based (red), classical (green), projection-based (black) and random (blue). Newly proposed methods (CAA, CRA, CAR, CPA, CJC, LCL, CN, AA, RA and JC) are marked by a white shine in their bar-plot and a "(N)" underneath the name of the respective method. Error bars represent the standard errors. The name, left and right node sizes of each network are reported above each evaluation plot (A-F).

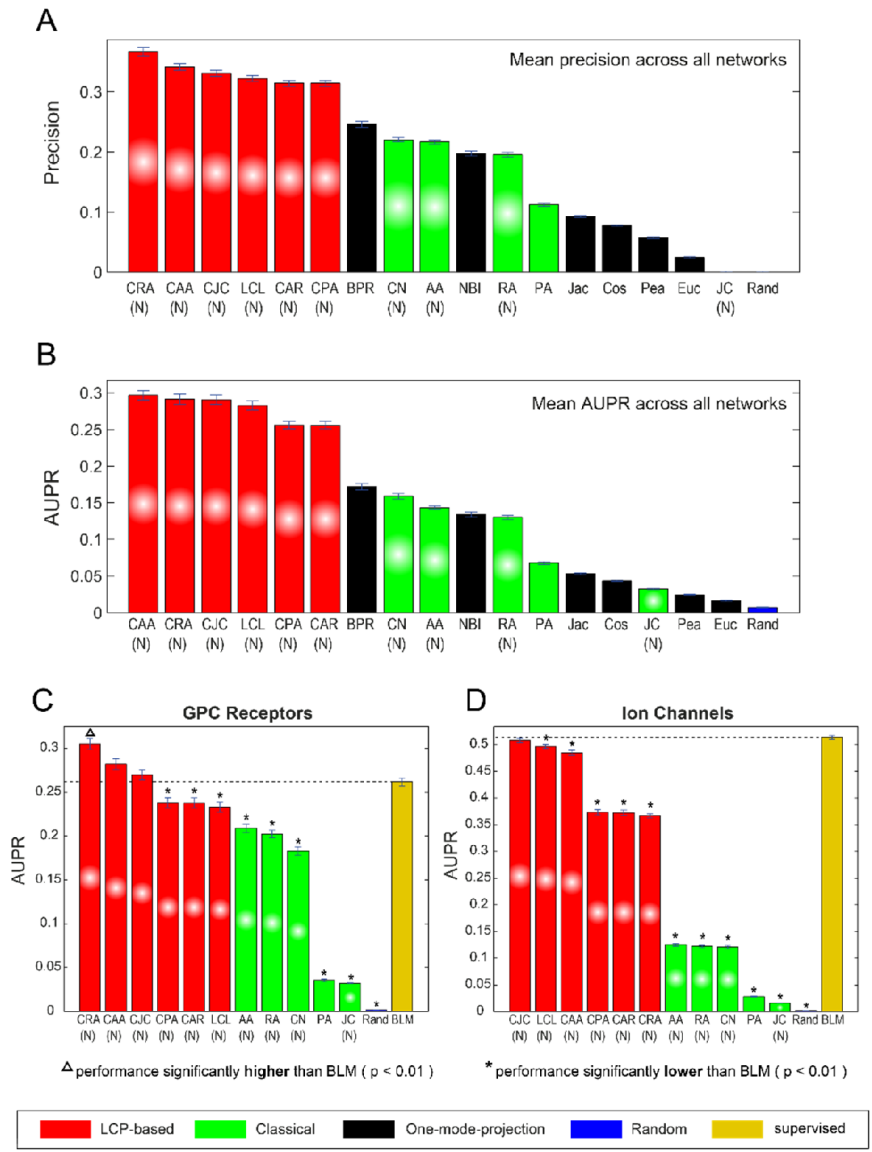

**Figure 3. Mean performance across all networks and comparison with the baseline supervised method. (A), (B)** Mean precision and mean area under the precision recall curve (AUPR) computed considering for each method the mean of 600 results obtained in 100 iterations in each of the 6 networks. Precision and AUPR in re-predicting 10% of randomly removed network links were measured. **(C), (D)** AUPR evaluates the ability of each method to rank the 10% removed network links the highest possible in the candidate list. Mean AUPR - computed for each network considering the mean of 100 iterations - is reported for two out of three biological networks. The name of each network is reported above each evaluation plot. The plot of the third network is reported in suppl. Fig. 2. P-values were calculated by the Mann-Whitney test and Benjamini-correction for multiple hypothesis testing was applied. Colours of the bars define the class of the prediction methods: LCP-based (red), classical (green), projection-based (black), random (blue) and supervised (yellow). Newly proposed methods (CAA, CRA, CAR, CPA, CJC, LCL, CN, AA, RA and JC) are marked by a white shine in their bar-plot and a "(N)" underneath the name of the respective method. Error bars represent the standard errors.

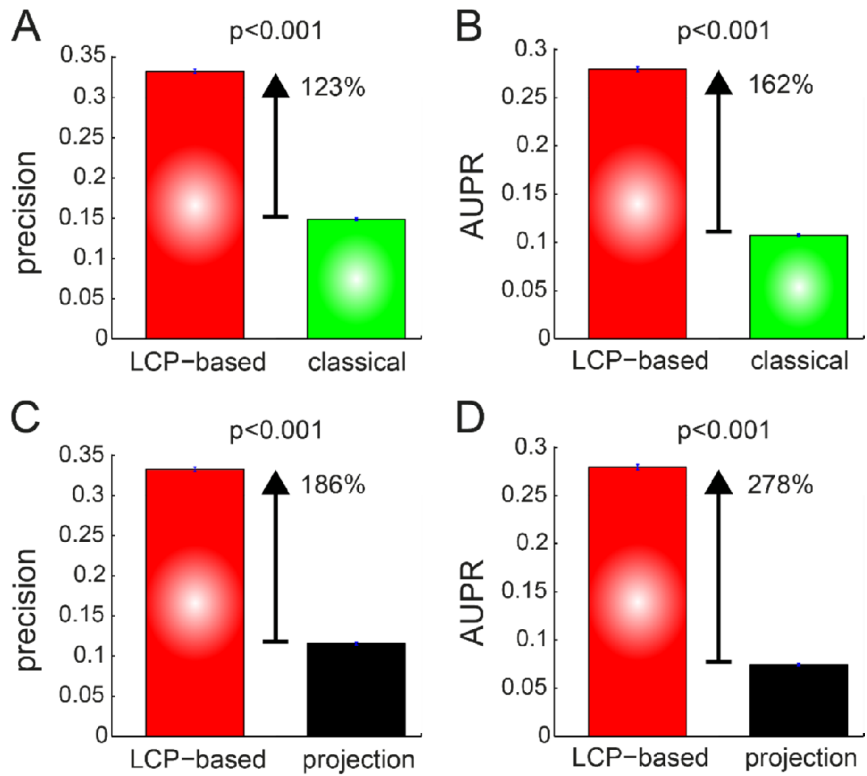

**Figure 4. Comparison of LCP-based and classical/projection methods. (A), (B)** Mean precision and mean AUPR for LCP-based (CRA, CAA, CAR, LCL, CPA, CJC) compared to classical (RA, AA, PA, JC, CN) methods. All LCP-based and classical methods (apart from PA) are proposed here for the first time in bipartite networks. These two newly proposed classes of bipartite models are marked by a white shine in their bar-plot. **(C), (D)** Mean precision and mean AUPR for LCP-based compared to one-mode-projection (BPR, NBI, Jac, Cos, Pea, Euc) methods. The mean for LCP-based class is obtained as the mean of 3600 results coming from 100 iterations of 6 methods applied in 6 networks. Following the same procedure, the mean of 3000 results for the classical class and the mean of 3600 results for the one-mode-projection class are reported. P-values were calculated with the Mann-Whitney test. Error bars represent the standard error and are negligible.

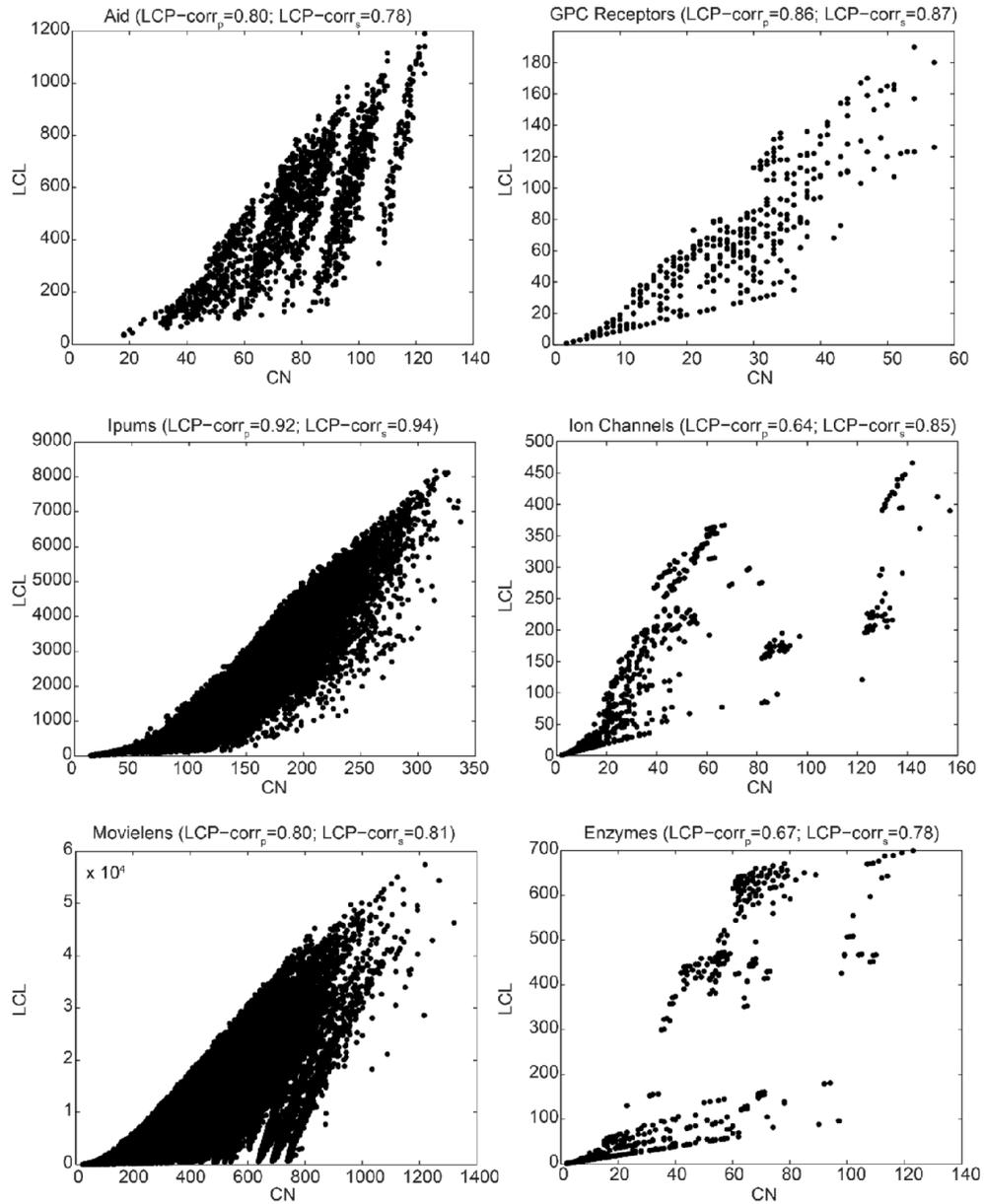

**Figure 5. Local-community-paradigm decomposition plot (LCP-DP) of the considered bipartite networks.** Each dot in the plot represents an existing link in the network. The x-axis is the number of common neighbours (CN) of the seed nodes of each existing link, while the y-axis is the number of the local community links (LCL) between the CNs. The title of each plot reports: the name of the network, the Pearson correlation value and finally the Spearman correlation value.

# SUPPLEMENTARY INFORMATION

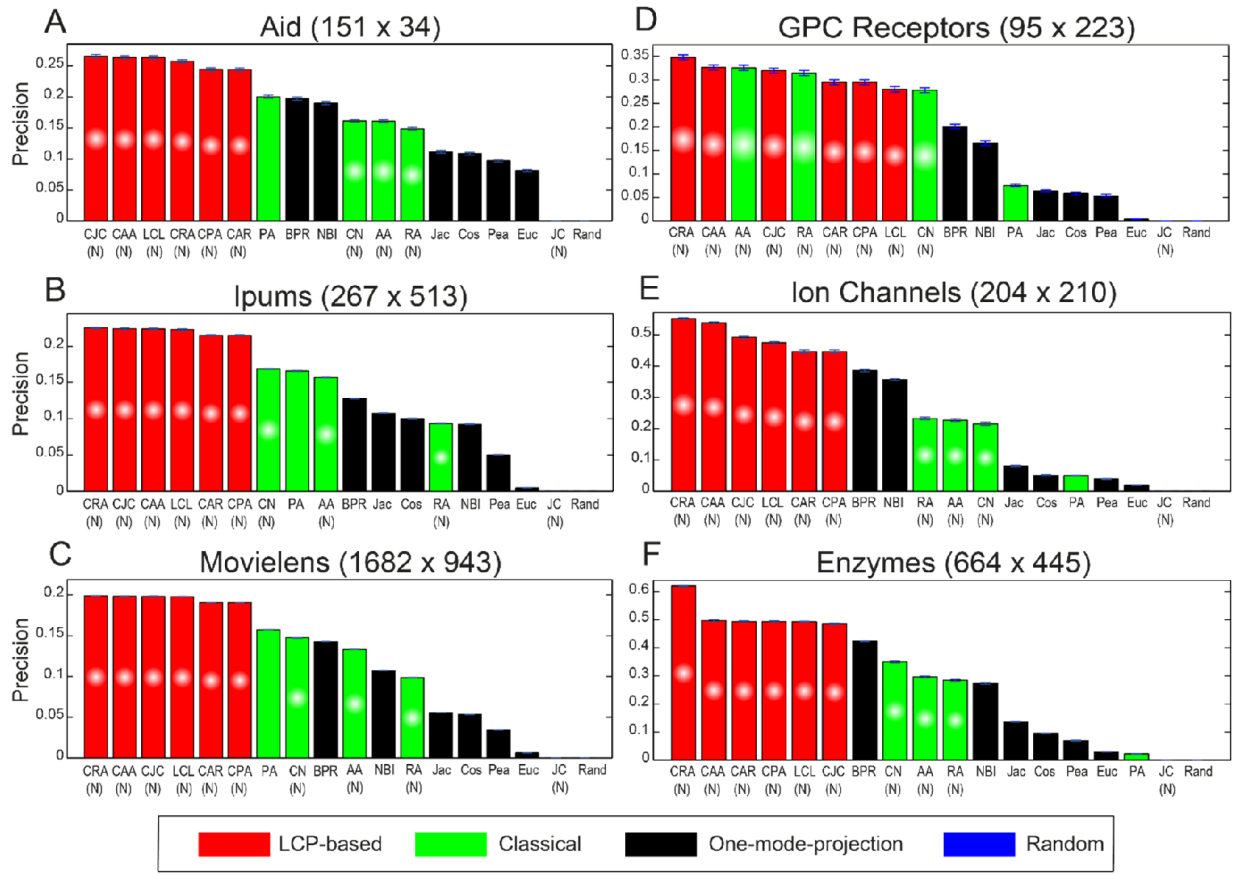

**Suppl. Fig. 1. Link prediction evaluation based on the models' precision.** The precision represents the ratio of correct edges recovered out of the 10% edges randomly removed from the original network. This operation was repeated 100 times for each network and the mean precision and standard error for each method are reported. Colours of the bars define the class of the prediction methods: LCP-based (red), classical (green), projection-based (black), random (blue). Newly proposed methods (CAA, CRA, CAR, CPA, CJC, LCL, CN, AA, RA and JC) are marked by a white shine in their bar-plot and a "(N)" underneath the name of the respective method. Error bars represent the standard errors. The name, left and right node sizes of each network are reported above each evaluation plot (A-F).

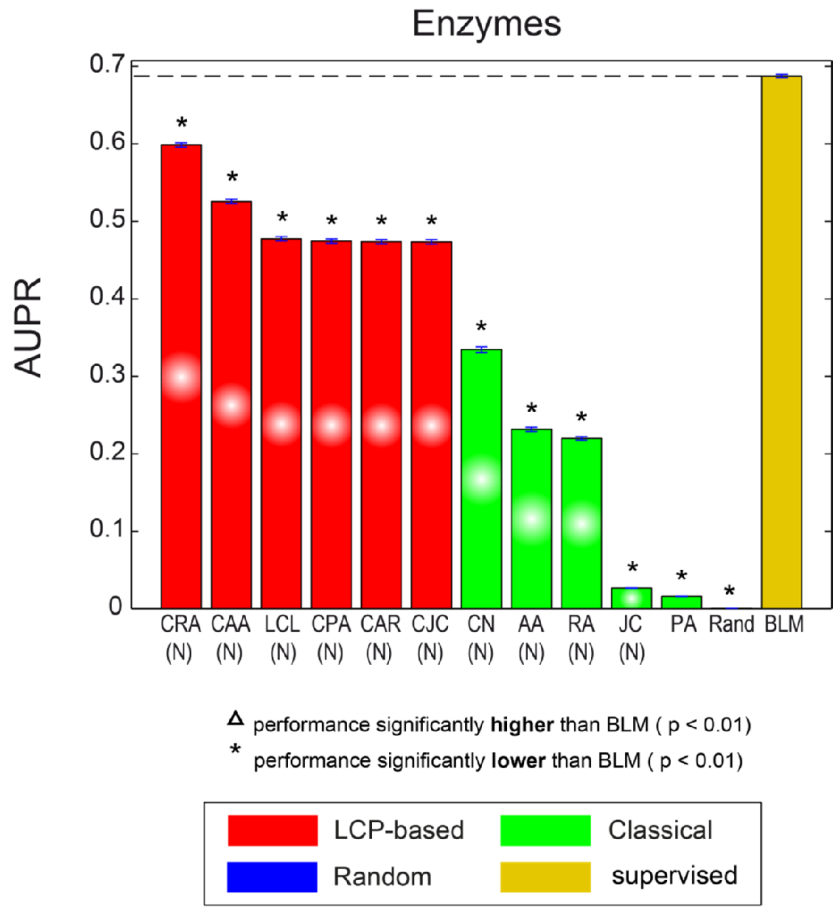

**Suppl. Fig. 2. Mean AUPR performance and comparison with the baseline supervised method in Enzymes biological network.** AUPR evaluates the ability of each method to rank the 10% removed network links the highest possible in the candidate list. Mean AUPR - computed for each network considering the mean of 100 iterations - is reported. P-values were calculated by the Mann-Whitney test and Benjamini-correction for multiple hypothesis testing was applied. Colours of the bars define the class of the prediction methods: LCP-based (red), classical (green), projection-based (black), random (blue) and supervised (yellow). Newly proposed methods (CAA, CRA, CAR, CPA, CJC, LCL, CN, AA, RA and JC) are marked by a white shine in their bar-plot and a "(N)" underneath the name of the respective method. Error bars represent the standard errors.

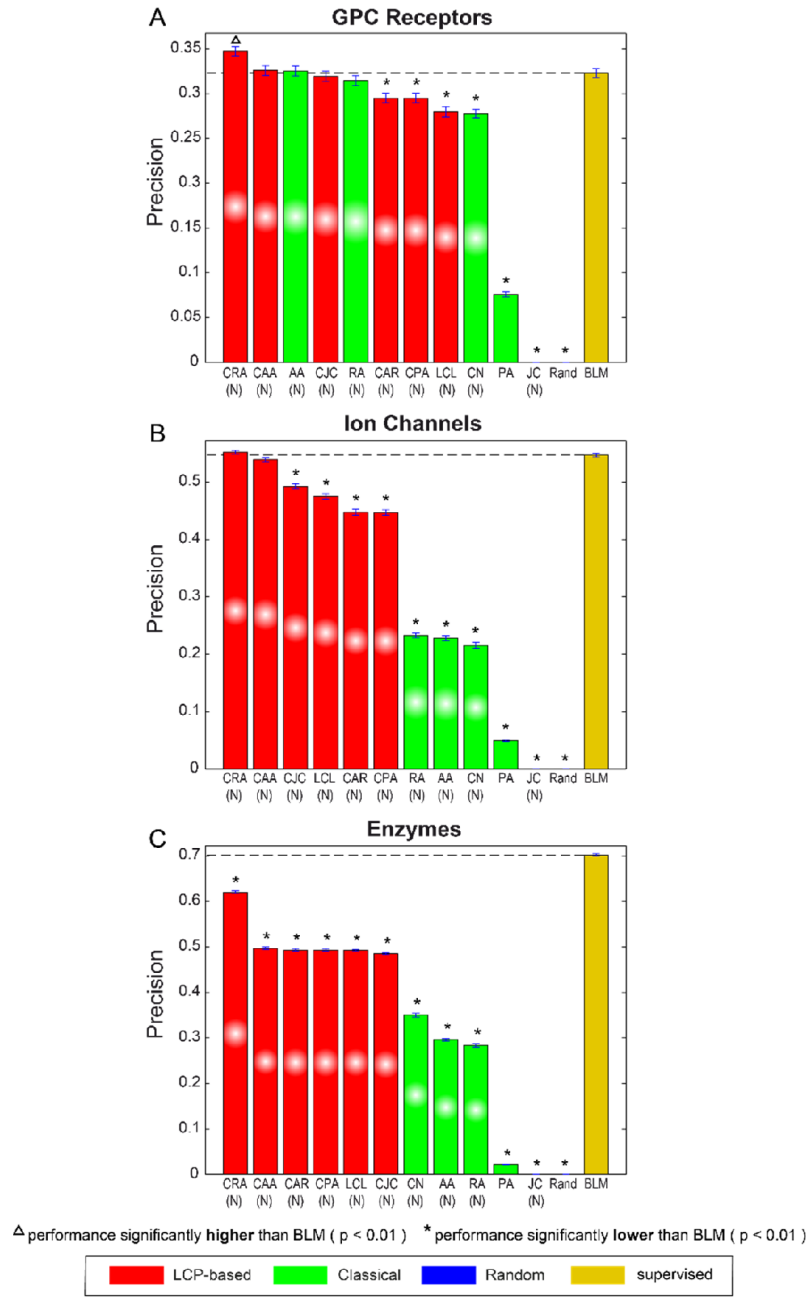

**Suppl. Fig. 3. Mean precision performance and comparison with the baseline supervised method in the three biological network.** Mean precision - computed for each network considering the mean of 100 iterations - is reported. P-values were calculated by the Mann-Whitney test and Benjamini-correction for multiple hypothesis testing was applied. Colours of the bars define the class of the prediction methods: LCP-based (red), classical (green), projection-based (black), random (blue) and supervised (yellow). Newly proposed methods (CAA, CRA, CAR, CPA, CJC, LCL, CN, AA, RA and JC) are marked by a white shine in their bar-plot and a "(N)" underneath the name of the respective method. Error bars represent the standard errors.

**Supplementary Table S1. Basic topological statistical features of the six used networks.**

| Network name | left average degree | right average degree | average degree | average Latapy clustering [1] | Robin Alexander clustering [2] | Betweenness centrality [3] | LCP Pearson/Spearman Correlation [4] |
|---|---|---|---|---|---|---|---|
| Aid | 12.51 | 55.56 | 10.21 | 0.2618 | 0.5011 | 0.0062 | 0.80/0.78 |
| Ipums | 67.75 | 35.26 | 23.19 | 0.1087 | 0.361 | 0.0021 | 0.92/0.94 |
| Movielens | 59.45 | 106.04 | 38.1 | 0.0715 | 0.2948 | 0.0006 | 0.80/0.81 |
| GPC Receptors | 2.85 | 6.68 | 2.0 | 0.4088 | 0.3637 | 0.006 | 0.86/0.87 |
| Ion channels | 7.03 | 7.24 | 3.57 | 0.3369 | 0.4814 | 0.0072 | 0.64/0.85 |
| Enzymes | 6.58 | 4.41 | 2.64 | 0.5813 | 0.7393 | 0.0028 | 0.67/0.78 |